\documentclass[journal, letterpaper, twoside, twocolumn]{IEEEtran} 

\newlength{\figwidth}

\usepackage{balance}

\usepackage[nosort]{cite}        
\usepackage{url} 
\usepackage[intlimits]{amsmath}
\usepackage{bbm}
\usepackage{graphicx}
\usepackage{paralist}
\usepackage{fancyref}
\usepackage[caption=false,font=footnotesize]{subfig}
\usepackage[stretch=16, shrink=16, step=4]{microtype}
\usepackage{vmr-symbols-vecbold}
\usepackage{standard-macros}

\usepackage{enumitem}
\usepackage{microtype}
\usepackage{booktabs}
\usepackage{xcolor}
\usepackage{setspace}
\usepackage[normalem]{ulem}

 \usepackage{color}
 \definecolor{links}{rgb}{0.7,0,0}   
 \definecolor{urls}{rgb}{0,0,0.8}    
 \definecolor{cites}{rgb}{0,0,0.8}   
 \usepackage[colorlinks,hyperindex,linkcolor=black,citecolor=black,urlcolor=black]{hyperref} 

\newcommand{\vecrp}{\vecr_{n}^{(p)}}
\newcommand{\vecyp}{\vecy_{n}^{(p)}}
\newcommand{\vecdp}{\vecd_{n}^{(p)}}
\newcommand{\vecwp}{\vecw_{n}^{(p)}}

\newcommand{\matXp}{\matX_p}

\IEEEoverridecommandlockouts

\begin{document}

\title{Throughput Analysis of Massive~MIMO Uplink with Low-Resolution~ADCs}
%
\author{Sven Jacobsson,~\IEEEmembership{Student Member,~IEEE,} Giuseppe~Durisi,~\IEEEmembership{Senior Member,~IEEE,} Mikael Coldrey,~\IEEEmembership{Member,~IEEE,} Ulf Gustavsson, and Christoph~Studer,~\IEEEmembership{Senior~Member,~IEEE}
\thanks{S.\ Jacobsson is with Ericsson Research and Chalmers University of Technology, Gothenburg, Sweden (e-mail: \url{sven.jacobsson@ericsson.com})}
\thanks{G.\ Durisi is with Chalmers University of Technology, Gothenburg, Sweden (e-mail: \url{durisi@chalmers.se})}
\thanks{M.\ Coldrey and U.\ Gustavsson are with Ericsson Research, Gothenburg, Sweden (e-mail: \url{{mikael.coldrey, ulf.gustavsson}@ericsson.com})}
\thanks{C.\ Studer is with Cornell University, Ithaca, NY (e-mail: \url{studer@cornell.edu})}
\thanks{The work of S.\ Jacobsson and G.\ Durisi was supported in part by the Swedish Foundation for Strategic Research under grant ID14-0022, and by the Swedish Government Agency for Innovation Systems (VINNOVA) within the competence center~ChaseOn.}
\thanks{The work of C. Studer was supported in part by Xilinx Inc., and by the US National Science Foundation (NSF) under grants ECCS-1408006, CCF-1535897, and CAREER CCF-1652065.} 
\thanks{The material in this paper was presented in part at the IEEE International Conference on Communications (ICC) Workshop on 5G and Beyond: Enabling Technologies and Applications, London, U.K., June 2015~\cite{jacobsson15a}.}
}

\maketitle


\begin{abstract}

We investigate the uplink throughput achievable by a multiple-user (MU) massive multiple-input multiple-output (MIMO) system in which the base station is equipped with a large number of low-resolution analog-to-digital converters (ADCs). 
Our focus is on the case where neither the transmitter nor the receiver have any \emph{a priori} channel state information.
This implies that the fading realizations have to be learned through pilot transmission followed by channel estimation at the receiver, based on coarsely quantized observations.
%
%
{We  propose a novel channel estimator, based on Bussgang's decomposition, and a novel approximation to the rate achievable with finite-resolution ADCs, both for the case of finite-cardinality constellations and of Gaussian inputs, that is accurate for a broad range of system parameters.}
{Through numerical results, we illustrate that, for the 1-bit quantized case, pilot-based channel estimation together with maximal-ratio combing or zero-forcing detection enables reliable multi-user communication with high-order constellations in spite of the severe nonlinearity introduced by the ADCs.} 
{Furthermore, we show that the rate achievable in the infinite-resolution (no quantization) case can be approached using ADCs with only a few bits of resolution. }
{We finally investigate the robustness of low-ADC-resolution MU-MIMO uplink against receive power imbalances between the different users, caused for example by imperfect power~control.}

\end{abstract}


\section{Introduction}

Massive multiple-input multiple-output (MIMO) is a promising multi-user (MU) MIMO technology for next generation cellular communication systems (5G) \cite{larsson14a}. With massive MIMO, the number of antennas at the base station (BS) is scaled up by several orders of magnitude compared to traditional multi-antenna systems with the goals of enabling significant gains in capacity and energy efficiency~\cite{larsson14a, marzetta10a}.
Increasing the number of BS antenna elements leads to high spatial resolution; this makes it possible to simultaneously serve several user equipments (UEs) in the same time-frequency resource, which brings large capacity gains.
The improvements in terms of radiated energy efficiency are enabled by the array gain that is provided by the large number of~antennas.
 
Equipping the BS with a large number of antenna elements, however,  increases considerably the hardware cost and the power consumption of the radio-frequency (RF) circuits~\cite{yang13a}. 
This calls for the use of low-cost and power-efficient hardware components, which, however, reduce the signal quality due to an increased level of impairments.
{The aggregate impact of hardware impairments on massive MIMO systems has been investigated in, e.g., \cite{gustavsson14a, zhang14a, bjornson14a, bjornson15c}, where it is found that massive MIMO provides---to a certain extent---robustness against signal distortions caused by low-cost RF components. 
}{However, most of these analyses rely on the assumption that the distortion caused by the hardware imperfections can be modeled as an additive Gaussian random variable that is independent of the transmit signal.  
It is \emph{prima facie} unclear how accurate such modeling assumption is, especially for the distortion caused by low-resolution analog-to-digital converters (ADCs). This has been noted in~\cite[Sec.~IV.A]{bjornson15c}  where it is pointed out that such modeling assumption targets ADCs with high resolution.
}


\subsection{Quantized Massive MIMO}

{In this paper, we consider an uplink massive MU-MIMO system and focus on the signal distortion caused by the use of low-resolution ADCs at the BS. }
An ADC with sampling rate~$f_s$\,Hz and a resolution of $b$\,bits maps each sample of the continuous-time, continuous-amplitude baseband received signal to one out of $2^b$ quantization labels, by operating $f_s \cdot 2^b$ conversion steps per second.
In modern high-speed ADCs (e.g., with sampling rates larger than $1$\,GS/s), the dissipated power scales exponentially in the number of  bits and linearly in the sampling rate \cite{walden99a, murmanna}. 
This implies that for wideband massive MIMO systems where hundreds of high-speed converters are required, the resolution of the ADCs may have to be kept low in order to maintain the power consumed at the BS within acceptable~levels.

An additional motivation for reducing the ADC resolution is to limit the amount of data that has to be transferred over the link that connects the RF components and the baseband-processing unit.
For example, consider a BS that is equipped with an antenna array of $500$ elements. At each antenna element, the in-phase and quadrature samples are quantized separately using a pair of $10$-bit ADCs operating at $1$\,GS/s. Such a system would produce $10$~Tbit/s of data. This exceeds by far the rate supported by the common public radio interface (CPRI) used over today's fiber-optical fronthaul links \cite{ericsson-ab13a}.
Alleviating this capacity bottleneck is of particular importance in a cloud radio access network (C-RAN) architecture \cite{park14a}, where the baseband processing is migrated from the BSs to a centralized unit, which may be placed at a significant distance from the~BS~antenna~array.

An implication of lowering the ADC resolution is that the requirement on accurate radio-frequency circuitry can be relaxed. The reason is that the quantization noise may dominate the noise introduced by other components such as mixers, oscillators, filters, and low-noise amplifiers. Hence, further power-consumption reductions may be achieved by relaxing the quality requirements on the RF circuitry. 

The 1-bit resolution case, where the in-phase and quadrature components of the continuous-valued received samples are quantized separately using a pair of 1-bit ADCs, is particularly attractive because of the resulting low hardware complexity~\cite{odonnell05a, hoyos05a}. Indeed, a 1-bit ADC can be realized using only a simple comparator. Furthermore, in a 1-bit architecture, there is no need for automatic gain control circuitry, which is otherwise needed to match the dynamic range of the~ADCs.

\subsection{Previous Work}

%
Receivers employing low-resolution ADCs need to cope with the severe nonlinearity introduced by the coarse quantization, which may render signaling schemes and receiver algorithms developed for the case of high-resolution ADCs~suboptimal.

The impact of the 1-bit ADC nonlinearity on the performance of communication systems has been previously studied in the literature under various channel-model assumptions. 
In \cite{singh09a}, it is proven that BPSK is capacity achieving over a real-valued nonfading single-input single-output (SISO) Gaussian channel.
For the complex-valued Gaussian channel, QPSK is optimal.

These results hold under the assumption that the 1-bit quantizer is a zero-threshold comparator. It turns out that in the low-SNR regime, a zero-threshold comparator is not optimal  \cite{koch13a}. The optimal strategy involves the use of \emph{flash-signaling}~\cite[Def.~2]{verdu02a} and requires an optimization over the threshold value.
Unfortunately, the power gain obtainable using this optimal strategy manifests itself only at extremely low values of spectral efficiency. 

For the Rayleigh-fading case, under the assumption that the receiver has access to perfect channel state information (CSI), it is shown in \cite{krone10a} that QPSK is capacity achieving (again for the SISO case).
The assumption that perfect CSI is available may, however, be unrealistic in the 1-bit quantized case, since the nonlinear distortion caused by the 1-bit ADCs makes channel estimation challenging.
In particular, if the fading process evolves rapidly, the cost of transmitting training symbols cannot be neglected.
For the more practically relevant case when the channel is not known \emph{a priori} to the receiver, but must be learned (for example, via pilot symbols), QPSK is optimal when the SNR exceeds a certain threshold that depends on the coherence time of the fading process~\cite{mezghani08a}. 
For SNR values that are below this threshold, on-off QPSK is capacity achieving~\cite{mezghani08a}.

For the 1-bit quantized MIMO case, the capacity-achieving distribution is unknown.
In \cite{mezghani07a}, it is shown that QPSK is optimal at low SNR, again under the assumption of perfect CSI at the receiver.
{Mo and Heath Jr.~\cite{mo15b} derived high-SNR bounds on capacity, and showed that high-order modulations are supported. 
However, their analysis relies on the assumption that the transmitter has access to perfect CSI, which is unrealistic in low-resolution architectures. Their contribution leaves open the question on whether high-order modulations are supported in training based schemes where the receiver has partial knowledge of the channel and the transmitter (in our case, the UE) has no channel knowledge.}

Channel estimation on the basis of quantized observations is considered in, e.g.,~\cite{lok98a, ivrlac07a} (see also \cite{zymnis10a} for a compressive-sensing version of this problem). A closed-form solution for the maximum likelihood (ML) estimate in the 1-bit case is derived in \cite{ivrlac07a}, under the assumption of time-multiplexed pilots.

The use of 1-bit ADCs in massive MIMO was considered in~\cite{risi14a}. There, the authors examined the achievable uplink throughput for the scenario where the UEs transmit QPSK symbols, and the BS employs a least squares (LS) channel estimator, followed by a maximal ratio combining (MRC) or zero-forcing (ZF) detector.
Their results show that large sum-rate throughputs can be achieved despite the coarse quantization.
The results in \cite{risi14a} were extended to high-order modulations (e.g., 16-QAM) by the authors of this paper in \cite{jacobsson15a}. 
{There, we showed that one can detect not only the phase, but also the amplitude of the transmitted signal, provided that the number of BS antennas is sufficiently large, hence, answering positively the question left open in~\cite{mo15b}.}
Choi~\emph{et al.} \cite{choi15a} recently developed a detector and a channel estimator capable of supporting high-order constellations such as 16-QAM.
Again for the case of 1-bit ADCs, Li \emph{et al.}~\cite{li16a,li17b} proposed a linear minimum mean square error (LMMSE) channel estimator based on Bussgang's decomposition that was shown to be superior to the one proposed in~\cite{choi15a}. 
Furthermore, they derive an approximation on the rates achievable with Gaussian inputs. The accuracy of this approximation is not fully validated in~\cite{li17b}, since no comparison with actual achievable rates is provided.
Wen~\emph{et al.}~\cite{wen15b}  proposed a joint channel- and data-estimation algorithm that offers significant improvement compared to the case when channel estimation and data detection are treated separately. However, as noted in \cite{wen15b}, the complexity of the proposed algorithm is too high for practical implementations.

A mixed-ADC architecture where many 1-bit ADCs are complemented with few high-resolution ADCs is proposed in \cite{liang15a}. It is found that the addition of a relatively small number of high-resolution ADCs increases the system performance significantly. 
{Specifically, the authors of~\cite{liang15a} present an achievability bound under Gaussian signaling and minimum distance decoding that holds for the setup where channel estimates are acquired through the high-resolution ADCs.
This relies on the assumption that each high-resolution ADC can be linked to several RF chains through a switch. The disadvantage of such architecture is that ADC switches increase hardware complexity. Furthermore, the time needed to acquire channel estimates  increases dramatically.}

In all of the contributions reviewed so far, low-resolution quantized massive MIMO systems have been investigated solely for communication over frequency-flat, narrowband, channels. 
A spatial-modulation-based massive MIMO system over a frequency-selective channel was studied in \cite{wang14b}. The proposed receiver employs LS estimation followed by a message-passing-based detector.
The performance of a low-resolution quantized massive MIMO system using orthogonal frequency division multiplexing (OFDM) and operating over a wideband channel was investigated in~\cite{studer16a}.
There, it is found that using ADCs with only $4$ to $6$ bits resolution is sufficient to achieve performance close to the infinite-resolution (i.e., no quantization) case, at no additional cost in terms of digital signal processing complexity.
{A capacity lower bound for wideband channels and 1-bit ADCs has been recently reported in~\cite{mollen16c}.
The analysis in~\cite{mollen16c} relies on the same signal decomposition used in~\cite{li16a,li17b} for the frequency-flat case.
However, differently from~\cite{li16a,li17b}, the temporal correlation of the quantization noise in the channel-estimation phase is~ignored.}

All  the results reviewed so far hold under the assumption of Nyquist-rate sampling at the receiver.
It is worth pointing out that Nyquist-rate sampling is not optimal in the presence of quantization at the receiver \cite{shamai94a, koch10a,zhang12a}.
For example, for the 1-bit quantized complex AWGN channel, high-order constellations such as 16-QAM can be supported even in the SISO case, if one allows for oversampling at the receiver \cite{krone12a}. 

\subsection{Contributions}
Focusing on Nyquist-rate sampling, and on the scenario where neither the transmitter nor the receiver have \emph{a priori} CSI, we investigate the rates achievable over a frequency-flat Rayleigh block-fading MU-MIMO channel, when the receiver is equipped with low-resolution ADCs.
Our contributions are summarized as follows:
\begin{itemize}  
\item {We propose a novel channel estimator for the case of multi-bit ADCs and nonuniform quantization regions using Bussgang's decomposition. 
This estimator recovers the LMMSE estimator  proposed in~\cite{li16a,li17b,mollen16c} for the case of 1-bit ADCs. }
\item {We present a easy-to-evaluate approximation on the rates achievable with finite-cardinality constellations under the assumption of training-based channel estimation.
The approximation is explicit in the number of pilots used to estimate the channel and in the resolution of the ADCs; by comparing it with a numerically computed lower bound on the achievable rates, we show that this approximation is accurate for a large range of SNR values.}
\item {We also obtain a closed-form approximation on the rates achievable with Gaussian inputs that is derived using Bussgang's decomposition. This  approximation recovers for the 1-bit case the approximation recently presented in~\cite{li16a,li17b}. A comparison with a numerically computed lower bound on the achievable rates reveals that, in the 1-bit case, this Gaussian approximation is  accurate at low SNR, but  overestimates the achievable rate at high SNR in the multiuser scenario.}
\item Through a numerical study, we determine the minimum ADC resolution needed to make the performance gap to the infinite-resolution case negligible. Our simulations suggest that only few bits (e.g., $3$ bits) are required to achieve a performance close to the infinite-resolution case for a large range of system parameters. This holds also when the users are received at vastly different power levels (imperfect power control).
\end{itemize}

This paper complements the analysis previously reported in~\cite{jacobsson15a} by generalizing it to ZF receivers, to multi-bit quantization, and to the case of imperfect power control. {Furthermore, the proposed channel estimator and the rate approximations are novel.}

\subsection{Notation}

Lowercase and uppercase boldface letters denote column vectors and matrices, respectively. 
The identity matrix of size $N \times N$ is denoted by $\matI_N$. 
{We use $\tr(\cdot)$ and $\diag(\cdot)$ to denote the trace and the main diagonal of a matrix}, and $\vecnorm{\cdot}$ to denote the~$\ell_2$-norm of a vector.
The multivariate normal distribution with mean $\vecmu$ and covariance $\matSigma$ is denoted by $\mathcal{N}(\vecmu,\matSigma)$.
Furthermore, the multivariate complex-valued circularly-symmetric Gaussian probability density function with zero mean and covariance~$\matSigma$ is denoted by $\mathcal{CN}(\veczero,\matSigma)$.
The operator $\Ex{x}{\cdot}$ stands for the expectation over the random variable $x$.
The mutual information between two random variables $x$ and $y$ is indicated by $I(x;y)$.
The real and imaginary parts of a complex scalar $s$ are $\Re\{s\}$ and $\Im\{s\}$.
The superscripts~$^T$, $^*$, and $^H$ denote transpose, complex conjugate, and Hermitian transpose, respectively. The function $\Phi(x)$ is the cumulative distribution function (CDF) of a standard normal random variable.

\subsection{Paper Outline}

The rest of the paper is organized as follows. 
In Section~\ref{sec:system_model}, we introduce the massive MIMO system model and the channel-estimation and data-detection problems. 
{In \fref{sec:rate_analysis}, we derive an approximation on the rate achievable with finite-cardinality constellations and Gaussian inputs. 
In \fref{sec:one_bit_num}, we validate the accuracy of our approximations for different scenarios and determine the ADC resolution required to approach the rate achievable in the infinite-resolution case. 
We conclude in Section~\ref{sec:conclusions}.}


\section{Channel Estimation and Data Detection with Low-Resolution ADCs}

\label{sec:system_model}

\subsection{System Model and Sum-Rate Capacity} 
\label{sec:system_model_and_sum_rate_capacity}

\begin{figure}[t]
	\centering
	\includegraphics[width = .85\figwidth]{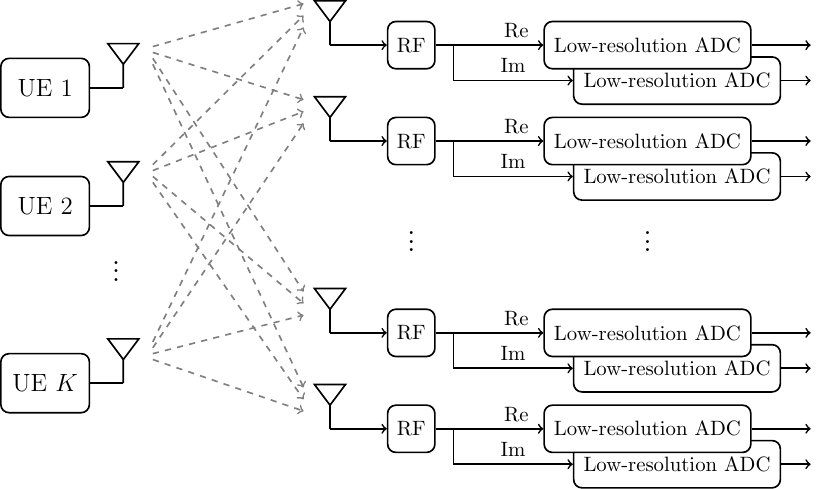}
	\caption{Quantized massive MIMO uplink system model.}
	\label{fig:system}
\end{figure}

We consider the single-cell uplink system depicted in Fig.~\ref{fig:system}. Here, $K$ single-antenna users are served by a BS that is equipped with an array of $N>K$ antennas. 
We model the subchannels between each transmit-receive antenna pair as a Rayleigh block-fading channel (see, e.g.,~\cite{marzetta99a}), i.e., a channel that stays constant for a block of $T$ channel uses, and changes independently from block to block. 
We shall refer to $T$ as the channel coherence interval. 
We further assume that the subchannels are mutually independent.
The discrete-time complex baseband received signal over all antennas within an arbitrary coherence interval and before quantization is modeled~as 
\begin{IEEEeqnarray}{rCL}
	\rvecy_t &=& \rmatH \vecx_t  + \rvecw_t, \quad t = 1, 2, \dots, T.
\end{IEEEeqnarray}
Here, $\vecx_t \in \opC^{K}$ denotes the channel input from all users at time~$t$, and $\rmatH \in \opC^{N\times K}$ is the channel matrix connecting the $K$ users to the $N$ BS antennas.
The entries of $\rmatH$ are independent and $\mathcal{CN}(0,1)$-distributed.
Furthermore, the vector $\rvecw_t \in \opC^{N}$, whose entries are  independent and $\mathcal{CN}(0,1)$-distributed, stands for the AWGN.

Throughout the paper, we consider the case where CSI is not available \emph{a priori} to the transmitter or to the receiver, i.e., they are both not aware of the realization of~$\rmatH$. 
This scenario captures the cost of learning the fading channel~\cite{lapidoth05a,durisi16a,yang13b}, an operation that has to be performed using quantized observations and may yield significant performance loss in the case of low-resolution ADCs.
We further assume that coding can be performed over many coherence intervals.
Let $\rmatX = [\rvecx_1, \rvecx_2, \dots, \rvecx_T]$ be the $K \times T$ matrix of transmitted signals within a coherence interval, and let $\rmatR = [\rvecr_1, \rvecr_2, \dots, \rvecr_T]$ be the corresponding $N \times T$ matrix of received quantized samples.
For a given quantization function, the ergodic sum-rate capacity~is~\cite{marzetta99a} 
\begin{IEEEeqnarray}{rCL}
\label{eq:sumratecap}
	C(\snr) 
	&=& \frac{1}{T} \sup I(\rmatX; \rmatR).
\end{IEEEeqnarray}
Here, the supremum is over all probability distributions on $\rmatX$ for which $\rmatX$ has independent rows and the following average power constraint is satisfied:
\begin{equation}
\label{eq:avgpow}
\Ex{}{\tr(\matX\matX^H)} \le K T \snr.
\end{equation}
Since the noise variance is normalized to one, we can think of $\snr$ as the SNR.
The sum-rate capacity in \eqref{eq:sumratecap} is, in general, not known in closed form, even in the infinite-resolution case, for which tight capacity bounds have been  reported recently~in~\cite{devassy15a}.

\subsection{Quantization of a Complex-Valued Vector}

The in-phase and quadrature components of the received signal at each antenna are quantized separately by an ADC of $b$-bit resolution. 
We characterize the ADC by a set of $2^b+1$ quantization thresholds $\setT_b = \{\tau_0, \tau_1, \dots, \tau_{2^b} \}$, such that $-\infty =\tau_0 < \tau_1 < \dots < \tau_{2^b} = \infty$, and a set of $2^b$ quantization labels $\setL_b = \{ \ell_0, \ell_1, \dots, \ell_{2^b-1} \}$ where $\ell_i \in (\tau_i, \tau_{i+1}]$.
Let $\setR_b = \setL_b \times \setL_b$.
We shall describe the joint operation of the $2N$ $b$-bit ADCs at the BS by the function $Q_b(\cdot): \opC^{N} \rightarrow  \setR_b^{N}$ that maps the received signal $\rvecy_t$ with entries $\{ y_{n, t}\}$ to the quantized output $\rvecr_t$ with entries $\{ r_{n, t}\}$ in the following way: if $\Re\{ y_{n, t} \} \in (\tau_k, \tau_{k+1}]$ and $\Im\{ y_{n, t} \} \in (\tau_l, \tau_{l+1}]$, then $r_{n, t} = \ell_k + j\ell_l$.
Using this convention, the quantized received signal can be written~as
\begin{IEEEeqnarray}{rCL}
	\rvecr_t &=& Q_b ( \vecy_t ) = Q_b \lefto( \rmatH \vecx_t+ \rvecw_t \right), \quad t = 1, 2, \dots, T.
	\label{eq:inoutMIMO}
\end{IEEEeqnarray}

{Finding the optimal quantization labels, i.e., the ones that minimize the mean square error (MSE) between the nonquantized received vector $\vecy_t$ and the quantized vector $\vecr_t$, requires one to determine the probability density function (PDF) of~$\vecy_t$.
}
{Note that such PDF depends on the choice of the input constellation $\vecx_t$. }
{Since adapting the quantization labels to the choice of the input constellation appears to be impractical from an implementation point of view, in this paper we shall consider the following suboptimal choice of the set of quantization labels $\setL_b$ and thresholds~$\setT_b$: 
we first approximate  each entry of the nonquantized channel output vector $\vecy_t$
by a complex Gaussian random variable with variance $K\snr + 1$ and then use the Lloyd-Max algorithm~\cite{max60a, lloyd82a} to find a set of labels $\tilde\setL_b=\{ \tilde{\ell_0}, \tilde{\ell_1}, \dots, \tilde{\ell}_{2^b-1} \}$ that minimize the mean square error between the nonquantized and the quantized signal.
}
Then, we rescale the labels such that the variance of each entry of $\vecr_t$ is $K\rho+1$.
Specifically, we multiply each label in the set $\tilde\setL_b$ by the factor
\begin{IEEEeqnarray}{rCL} \label{eq:alpha}
\alpha = \sqrt{\frac{K\snr + 1}{2\sum\limits_{i = 0}^{2^b-1} \tilde\ell_i^2 \lefto( \Phi\lefto( \sqrt{\frac{2\tau_{i+1}^2}{K\snr + 1}} \right) - \Phi\lefto( \sqrt{\frac{2\tau_{i}^2}{K\snr + 1}} \right) \right)}} \IEEEeqnarraynumspace
\end{IEEEeqnarray}
to produce the set of labels $\setL_b = \alpha\tilde\setL_b$.

{Some comments on our choice are in order.}
{The Gaussian approximation is accurate at low SNR or when the number of UE is sufficiently large. When such conditions are not fulfilled, it may result in a suboptimal choice of the quantization labels.}
{The rescaling of the labels by $\alpha$ in~\eqref{eq:alpha} turns out to simplify the performance analysis (see Sections~\ref{sec:channel_est} and~\ref{sec:gaussapprox})}.

In the 1-bit case, we can write the quantized received signal at the $n$th antenna, at discrete time $t$, as follows: 
\begin{IEEEeqnarray}{rCL}
  Q_1(y_{n,t}) &=& \sqrt{\frac{K\snr + 1}{2}}\lefto(\text{sgn}\lefto(\Re\{ y_{n,t} \}\right) + j \text{sgn}\lefto(\Im\{ y_{n,t} \}\right)\right). \IEEEeqnarraynumspace
\end{IEEEeqnarray}
Here, $\text{sgn}(\cdot)$ denotes the signum function defined as
\begin{IEEEeqnarray}{rCL}
	 \text{sgn}(x) &=& 
	 \begin{cases}
	 	-1, &\text{if} \ x < 0 \\
	 	+1, &\text{if} \ x \ge 0.
	 \end{cases}
\end{IEEEeqnarray}

\subsection{Signal Decomposition using Bussgang's Theorem}

The quantization of a vector using finite-resolution ADCs causes a distortion that is correlated with the input to the quantizer. When the  input to the quantizer is Gaussian, we can use Bussgang's theorem \cite{bussgang52a} to decompose the quantized signal in the convenient form detailed in the following~theorem.

\begin{thm} \label{thm:general} {Let $\vecr = Q_b(\vecy)$ denote the output from a set of ADCs described by the set of labels~$\setL_b$ and the set of thresholds~$\setT_b$. Assume that~$\vecy \sim \jpg(\veczero_N, \matK)$ where $\matK \in \opC^{N \times N}$. Then, the quantized vector~$\vecr$ can be decomposed as 
}
\begin{IEEEeqnarray}{rCl} \label{eq:r_decomp}
{\vecr = \matG_b\vecy+ \vecd}
\end{IEEEeqnarray}
{where the quantization distortion $\vecd$ and $\vecy$ are uncorrelated. Furthermore, $\matG_b \in \opR^{N \times N}$ is the following diagonal matrix:\footnote{We use the convention that the function $\exp(\cdot)$ applied to a diagonal matrix acts element-wise on its diagonal entries.} 
}%
{\begin{IEEEeqnarray}{rCl} \label{eq:gainmatrix_general}
\matG_b &=& \diag\lefto(\matK\right)^{-{1}/{2}} \sum_{i = 0}^{2^b-1} \frac{\ell_i}{\sqrt{\pi}} \bigg(\!\exp\lefto( - \tau_i^2\diag\lefto(\matK\right)^{-1}\right) \nonumber\\
&& - \exp\lefto( - \tau_{i+1}^2\diag\lefto(\matK\right)^{-1}\right) \!\bigg). \IEEEeqnarraynumspace
\end{IEEEeqnarray}}%
{Here, $\ell_i$ corresponds to the $i$th element of the set of labels $\setL_b$  and $\tau_i$ to the $i$th element of the set of thresholds $\setT_b$.
}\end{thm}
\begin{IEEEproof}
 { See~\fref{app:appA}.}
\end{IEEEproof}
{Bussgang's theorem has been used previously in the literature to decompose the quantized signal in the 1-bit-ADC case (see, e.g., \cite{li16a, li17b}).
A generalization of this result to the case of multi-bit \emph{uniform} ADCs has been recently proposed in~\cite{jacobsson16c} in the context of downlink precoding. 
The more general result in~\fref{thm:general} allows one to handle the case of nonuniform quantizers as well.
}%
{For the special case  when $\diag(\matK) = (K \snr + 1)\matI_N$, which will turn out relevant for our analysis in \fref{sec:channel_est}, the matrix $\matG_b$ in~\eqref{eq:gainmatrix_general} reduces to
}%
\begin{IEEEeqnarray}{rCl}\label{eq:scaled_identity}
{  \matG_b = G_b\matI_N}
\end{IEEEeqnarray}%
with
\begin{IEEEeqnarray}{rCl}
  G_b &=& \sum_{i = 0}^{2^b-1} \frac{\ell_i}{\sqrt{\pi(K\snr +1)}} \lefto( e^{ - \frac{\tau_i^2}{K\snr +1}} - e^{ - \frac{\tau_{i+1}^2}{K\snr +1}} \right). \label{eq:Gb_glass}
\end{IEEEeqnarray}
Note that in the infinite-resolution case ($b=\infty$),  it follows from~\eqref{eq:r_decomp} that $\matG_\infty = \matI_N$ and, hence,~$G_\infty = 1$ (see~\eqref{eq:scaled_identity}). For the 1-bit-ADCs case, we have that $G_1 = \sqrt{2/\pi}$, a well-known result used recently in~\cite{li17b,mollen16c} to analyze the throughput achievable with 1-bit ADCs. 
We shall use the Bussgang decomposition to develop a channel estimator in the next section as well as an approximation on the rates achievable with Gaussian inputs in Section~\ref{sec:gaussapprox}.

\subsection{Channel Estimation}\label{sec:channel_est}
A common approach to transmitting information over fading channels whose realizations are not known \emph{a priori} to the receiver is to reserve a certain number of time slots in each coherence interval for the transmission of pilot symbols.
These pilots are then used at the receiver to estimate the fading channel.
Assume that $P$ pilot symbols are used in each coherence interval ($K \le P \le T$). 
We shall assume that the pilot sequences used by different UEs are pairwise orthogonal, i.e., that 
\begin{IEEEeqnarray}{rCl}\label{eq:orth_pilots}
\sum_{t=1}^P \vecx_t\vecx_t^H = P\snr\,\matI_K.
\end{IEEEeqnarray}
Let $\vech_{n}$ denote the channel vector whose entries contain the channel gain between the $k$th UE, $k=1,\dots, K$, and the $n$th BS antenna. Furthermore, let $\matXp = [\vecx_1, \ldots, \vecx_P]^T$ denote the matrix containing the $P$ pilot symbols transmitted by the $K$ UEs. Finally, let $\vecyp =  \matXp\vech_n + \vecwp$ and $\vecrp = Q_b(\vecyp)$ denote the nonquantized and quantized pilot sequences received at the $n$th antenna during the training phase.
Here,~$\vecwp = [w_{n,1}, \ldots, w_{n,P}]^T$ is the additive noise. 
For the 1-bit case, the LMMSE estimator of $\vech_n$ was obtained in~\cite{li17b}.
Proceeding similarly to~\cite{li17b}, we generalize the LMMSE estimator~\cite{li17b} to the multi-bit case.
{Specifically, let $\matC_{\vecyp}$ and $\matC_{\vecrp}$ be the covariance matrices of $\vecyp$~and~$\vecrp$, respectively.  
Using Bussgang's decomposition~\eqref{eq:r_decomp} (recall that both additive noise and fading are  Gaussian) and the fact that  $\diag(\matC_{\vecyp}) = (K \snr + 1)\matI_P$, which follows from~\eqref{eq:orth_pilots} and implies that $\matG_b = G_b\matI_P$ (see~\eqref{eq:gainmatrix_general}),
we conclude that the LMMSE estimator for the multi-bit case is
\begin{IEEEeqnarray}{rCl}
\hat\vech_{n} &=& 	G_b\,\matXp^H \matC_{\vecrp}^{-1} \vecrp. \label{eq:siso_est}
\end{IEEEeqnarray}
The computation of~\eqref{eq:siso_est} requires knowledge of the covariance matrix~$\matC_{\vecrp}$. For the case of 1-bit ADCs, one can compute~$\matC_{\vecrp}$ in closed form, as shown in \cite{li17b}. 
For the multi-bit case, however, $\matC_{\vecrp}$ is not known in closed form. 
To overcome this issue, we shall next present an alternative channel estimator, which is an approximation of~\eqref{eq:siso_est}, but admits a simple closed-form expression.
Using Bussgang's decomposition~\eqref{eq:r_decomp}, we write~$\matC_{\vecrp}$~as
\begin{IEEEeqnarray}{rCl}
\matC_{\vecrp} \! &=& G_b^2 \matC_{\vecyp} \! + \matC_{\vecdp} = G_b^2\matXp\matXp^H \! + G_b^2\matI_P \! + \matC_{\vecdp}. \,\IEEEeqnarraynumspace
\label{eq:expansion_of_corr_r}
\end{IEEEeqnarray}
Here, $\matC_{\vecdp}$ denotes the covariance matrix of the quantization distortion.
To simplify~\eqref{eq:expansion_of_corr_r}, we shall next assume that the off-diagonal elements of $\matC_{\vecdp}$ are zero, i.e., we shall ignore the temporal correlation of the quantizaton distortion. Specifically, we assume that
\begin{IEEEeqnarray}{rCL}\label{eq:approx_off_diag}
   \matC_{\vecdp}=(1-G_b^2)(K\snr+1)\matI_P.
\end{IEEEeqnarray}
The assumption in~\eqref{eq:approx_off_diag} is accurate in the low-SNR regime or when the number of UEs $K$ is large, and it is actually exact if the number of pilots $P$ coincides with the number of UEs $K$.
The  constant on the right-hand side of~\eqref{eq:approx_off_diag} follows from the power normalization~\eqref{eq:alpha}. 
Substituting~\eqref{eq:approx_off_diag} into~\eqref{eq:expansion_of_corr_r} and~\eqref{eq:expansion_of_corr_r} into~\eqref{eq:siso_est}, we obtain
\begin{IEEEeqnarray}{rCl}
\hat{\vech}_{n}
&=& G_b\matXp^H  \lefto( G_b^2\matXp\matXp^H  \right.\nonumber\\ && + \left. \lefto(G_b^2 + (1 - G_b^2)(K\snr + 1)\right)\matI_P \right)^{-1} \vecrp \\
&=& G_b\lefto( G_b^2\matXp^H\matXp \right.\nonumber\\ && + \left.  \lefto(G_b^2 + (1-G_b^2)(K\snr+1)\right) \matI_K \right)^{-1} \matXp^H \vecrp  \IEEEeqnarraynumspace \\
&=& G_b \lefto(G_b^2P\snr \right. \nonumber \\ && + \left. G_b^2 + (1-G_b^2)(K\snr+1)\right)^{-1} \!\matXp^H \vecrp.  \label{eq:siso_est_simpler}
\end{IEEEeqnarray}
Rewriting~\eqref{eq:siso_est_simpler} in matrix form, we obtain the following simplified estimator, which we shall use in the remainder of the paper:
\begin{IEEEeqnarray}{rCL}
  \hat{\matH} 
  &=& \frac{G_b\sum_{t=1}^P \vecr_t\vecx_t^H}{G_b^2P\snr + G_b^2 + (1-G_b^2)(K\snr+1)} . \IEEEeqnarraynumspace \label{eq:LS_MIMO}
\end{IEEEeqnarray}

Some comments on~\eqref{eq:LS_MIMO} are in order.
For the case of 1-bit ADCs, the estimator~\eqref{eq:LS_MIMO}  coincides with the one derived in~\cite{mollen16c}. 
Under the assumption that the number of pilots $P$ is equal to the number of UEs $K$, the covariance matrix $\matC_{\vecdp}$ is indeed diagonal, and the estimator~\eqref{eq:LS_MIMO} is actually the LMMSE estimator~\eqref{eq:siso_est}.
This fact has been observed in~\cite{li17b} for the 1-bit case.
For the infinite-resolution case ($G_{\infty} = 1$),~\eqref{eq:LS_MIMO} coincides with the classic minimum mean square error (MMSE) estimator (see, e.g.,~\cite{hassibi03a}).
Let $\matH = \hat\matH+\tilde\matH$ where $\tilde\matH$ denotes the estimation error.
Under the assumption that~\eqref{eq:approx_off_diag} holds,  the variance of the channel estimate and of the estimation error take the following forms:
\begin{IEEEeqnarray}{rCl}
\hat\sigma^2 &=& \frac{1}{NK}\Ex{}{\tr\lefto(\hat\matH\hat\matH^H\right)} \\
&=& \frac{G_b^2P\snr}{G_b^2P\snr + G_b^2 + (1-G_b^2)(K\snr + 1)} \label{eq:sigmahat}
\end{IEEEeqnarray}
and
\begin{IEEEeqnarray}{rCl}
\tilde\sigma^2 &=&   \frac{1}{NK}\Ex{}{\tr\lefto(\tilde\matH\tilde\matH^H\right)} \\
&=& \frac{G_b^2 + (1-G_b^2)(K\snr + 1)}{G_b^2P\snr + G_b^2 + (1-G_b^2)(K\snr + 1)}. \label{eq:sigmatilde}
\end{IEEEeqnarray}
For the case $P=K$, \eqref{eq:sigmahat} and \eqref{eq:sigmatilde} are exact.  
Note that in the infinite-resolution case ($G_\infty = 1$), \eqref{eq:sigmahat} and \eqref{eq:sigmatilde} recover well-known results for MMSE estimation~(see,~e.g.,~\cite[Eq.~(19)]{hassibi03a}).
}

\subsection{Data Detection}

We shall focus on the practically relevant case when the BS employs a linear receiver.
Linear receiver processing---although inferior to nonlinear processing techniques such as successive interference cancellation---is less computationally demanding and has been shown to yield good performance if the number of antennas exceeds significantly the number of active users~\cite{bjornson15a}.
We shall consider two types of linear receivers, namely MRC and ZF. 
Using either of the two methods, a soft estimate~$\hat{x}_{k,t}$ of the transmitted symbol $x_{k,t}$ from the $k$th user at time $t = P+1, P+2, \dots, T$ is obtained as follows:
\begin{IEEEeqnarray}{rCl}
\label{eq:linreceived}
\hat{x}_{k,t}&=& \veca_k^H \vecr_t.
\end{IEEEeqnarray}
Here, $\veca_k \in \opC^N$ denotes the receive filter for the $k$th user, which is given by
\begin{IEEEeqnarray}{rCl} \label{eq:linreceiverdef}
	\veca_k &=&
	\begin{cases}
		\hat{\vech}_k / \vecnorm{\hat{\vech}_k}^2 , & \text{for MRC} \\
		(\hat{\matH}^\dagger)_k, & \text{for ZF}
	\end{cases}
\end{IEEEeqnarray}
where $(\hat{\matH}^\dagger)_k$ is the $k$th column of the pseudo-inverse of the channel estimate matrix $\hat{\matH}^\dagger = \hat{\matH} (\hat{\matH}^H \hat{\matH} )^{-1}$.

\subsection{High-Order Modulation Formats with 1-bit ADCs: Why Does it Work?}
\label{sec:one_bit}

\begin{figure}[!t]
	\captionsetup[subfloat]{farskip=0pt}
	\centering
	\subfloat[$N=20$ antennas, $\snr = 0$\,dB.]{\includegraphics[width = .45\columnwidth]{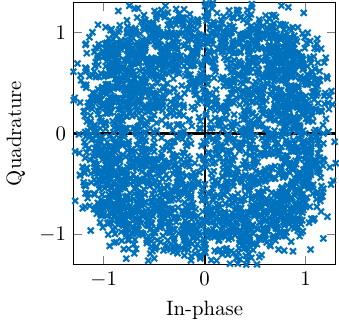}\label{fig:consta}} \quad
	\subfloat[$N=200$ antennas, $\snr = 0$\,dB.]{\includegraphics[width = .45\columnwidth]{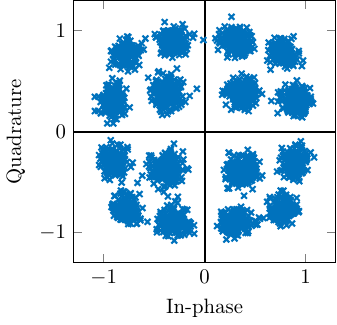}\label{fig:constb}} \\ \vspace{.3cm}
	\subfloat[$N=200$ antennas, $\snr = 20$\,dB.]{\includegraphics[width = .45\columnwidth]{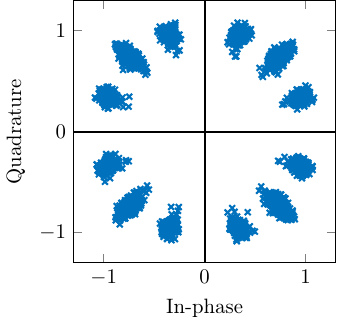}\label{fig:constc}}
	\caption{Single-user MRC outputs for 16-QAM inputs as a function of the number of receive antennas $N$ and the SNR $\snr$. The channel estimates are based on $P = 20$ pilot symbols.}
	\label{fig:const}
\end{figure}

Although for 1-bit ADCs, QPSK is optimal in the SISO case~\cite{mezghani08a}, the use of multiple antennas at the receiver opens up the possibility of using higher-order modulation schemes to support higher rates.
This observation is demonstrated in \fref{fig:const} where we plot the MRC receiver output (for 300 different channel fading realizations) for the case when a single user transmits 16-QAM data symbols.
The channel estimate is acquired using $P=20$ pilots.
As the size of the BS antenna array increases, the 16-QAM constellation becomes distinguishable (see \fref{fig:constb}), provided that $\snr$ is not too high.
Indeed, additive noise is one of the factors that enables the detection of the 16-QAM constellation. 
The other is the different phase of the fading coefficients associated with each receive antenna.
The explanation is as follows: in the 1-bit ADCs case, the quantized received output at each antenna belongs to the set $\setR_1$ of cardinality~$4$.
These $4$ possible outputs are then averaged by the MRC filter to produce an output (a scalar) that belongs to an alphabet with much higher cardinality.
The cardinality depends on the number of pilots and on the number of receive antennas.
The key observation is that the inner points of the 16-QAM constellation, which are more susceptible to noise, are more likely to be erroneously detected at each antenna. This results in a smaller averaged value after MRC than for the outer constellation points.

To highlight the importance of the additive noise, we consider in \fref{fig:constc} the case when $\snr = 20$\,dB. 
Since the additive noise is negligible, the output of the MRC filter lies approximately on a circle, which suggests that the amplitude of the transmitted signal cannot be used to convey information.
However, the phase of the 16-QAM symbols can still be detected. 
{Indeed, consider the following argument.
At high SNR and in the single-user case, the signal received at the $n$th antenna can be well-approximated by\footnote{In the remainder of this section, we shall drop the time index $t$ and the user index $k$ because they are superfluous.}
\begin{equation}\label{eq:high_snr_received_signal}
    r_n=Q_1(h_nx+w_n)\approx Q_1(h_nx)=Q_1(e^{j(\phi_n+\theta)}).
\end{equation}
Here, $\phi_n$ and $\theta$ denote the phase of $h_n$ and of $x$, respectively.
Furthermore, again at high SNR, the $n$th entry $a_n$ of the MRC filter $\veca$ in~\eqref{eq:linreceiverdef}
is well-approximated by
\begin{equation}\label{eq:high_snr_channel_estimate}
  a_n\approx \frac{1}{2N}Q_1(h_n)=\frac{1}{2N} Q_1\bigl(e^{j\phi_n}\bigr).
\end{equation}
Using~\eqref{eq:high_snr_received_signal} and~\eqref{eq:high_snr_channel_estimate}, we can approximate the MRC output~\eqref{eq:linreceived} at high SNR by
\begin{equation}\label{eq:high_snr_MRC_output}
  \hat{x}\approx\frac{1}{2N}\sum_{n=1}^N Q_1\bigl(e^{-j\phi_n}\bigr)Q_1\bigl(e^{j(\phi_n+\theta)}\bigr).
\end{equation}
To analyze~\eqref{eq:high_snr_MRC_output}, let us assume without loss of generality that $0<\theta<\pi/2$. 
Since $\phi_n$ is uniformly distributed~on $[0,2\pi]$ (recall that we assumed $h_n$ to be Rayleigh distributed), one can show that the phase of the random variable  $Q_1\bigl(e^{-j\phi_n}\bigr)Q_1\bigl(e^{j(\phi_n+\theta)}\bigr)$ is equal to $0$ with probability~$1-2\theta/\pi$ and is equal $\pi/2$ with probability $2\theta/\pi$. Hence, its mean is $\theta$. 
Since the fading coefficients $\{h_n\}$, and, hence, also their phases $\{\phi_n\}$, are independent, the phase of $\hat{x}$ in~\eqref{eq:high_snr_MRC_output} converges to $\theta$ as $N$ grows large, due to the central~limit~theorem. 

As shown in Fig.~\ref{fig:constc}, $N=200$ antennas are sufficient to distinguish the phase of  16-QAM constellation points at $20$ dB of SNR.
Note that independence between the $\{h_n\}$ is crucial for the central limit theorem to hold and for the phases to be distinguishable.}

\section{Achievable Rate Analysis}
\label{sec:rate_analysis}

{In this section, we shall characterize the rate achievable in a low-resolution quantized massive MIMO uplink system. In contrast to \cite{li16a, li17b, zhang15c, fan15a} we shall mainly focus on finite-cardinality constellations. Using Bussgang's decomposition, we also provide a closed-form approximation of the achievable rate with Gaussian inputs, which turns out accurate at low~SNR.}

\subsection{Sum-Rate Lower-Bound for Finite-Cardinality Inputs}

It follows from, e.g., \cite{tong04a}, that the achievable rate $R^{(k)}(\snr)$ for user $k = 1, 2, \dots, K$ with pilot-based channel estimation and MRC or ZF detection is
\begin{IEEEeqnarray}{rCl} \label{eq:achRate_MassiveMIMO}
R^{(k)}(\snr) &=& \frac{T-P}{T} I(x_k; \hat{x}_k \given \hat{\matH})
\end{IEEEeqnarray}
where $x_k$ and $\hat{x}_k$ are distributed as $x_{k,t}$  and $\hat{x}_{k,t}$ respectively.
It follows that the sum-rate capacity can be lower-bounded as follows:
\begin{IEEEeqnarray}{rCL}
	C(\snr)\geq \sum_{k=1}^K R^{(k)}(\snr).
\end{IEEEeqnarray}
In order to compute the achievable rate, we expand the mutual information in \eqref{eq:achRate_MassiveMIMO} as follows:
\begin{IEEEeqnarray}{rCl}
I(x_k; \hat{x}_k \given \hat{\matH}) 
&=& \Ex{x_k, \hat{x}_k, \hat{\matH}}{\log_2 \frac{P_{\hat{x}_k | x_k, \hat{\matH}}(\hat{x}_k | x_k, \hat{\matH})}{P_{\hat{x}_k | \hat{\matH}}(\hat{x}_k | \hat{\matH})}}. \IEEEeqnarraynumspace
\label{eq:mutinfo_MassiveMIMO}
\end{IEEEeqnarray}
To compute~\eqref{eq:mutinfo_MassiveMIMO}, one needs the conditional probability mass functions $P_{\hat{x}_k | x_k, \hat{\matH}}(\hat{x}_k | x_k, \hat{\matH})$ and $P_{\hat{x}_k | \hat{\matH}}(\hat{x}_k | \hat{\matH}) = \Ex{x_k}{P_{\hat{x}_k | x_k, \hat{\matH}}(\hat{x}_k | x_k, \hat{\matH})}$.
Since no closed-form expressions are available for these quantities, we  estimate them by Monte-Carlo sampling. 
Specifically, we simulate many noise and interference realizations, and map the resulting $\hat{x}_k$ to points over a rectangular grid in the complex plane. 
With this technique, one obtains a lower bound on $R^{(k)}(\snr)$~\cite[p.~3503]{arnold06a} that becomes increasingly tight as the grid spacing is made smaller.\footnote{The numerical routines used to evaluate~\eqref{eq:achRate_MassiveMIMO} can be downloaded at \url{https://github.com/infotheorychalmers/1-bit_massive_MIMO}.}
Note that \eqref{eq:mutinfo_MassiveMIMO} holds for every choice of input distribution and for ADCs with arbitrary resolution.

\subsection{Sum-Rate Approximation for Finite-Cardinality Inputs}

The evaluation of~\eqref{eq:mutinfo_MassiveMIMO} using the method just described is extremely time consuming.
{We next provide an accurate approximation of~\eqref{eq:mutinfo_MassiveMIMO} for finite-cardinality constellations that is easier  to evaluate, although still not in closed form} {(note that even for the infinite-resolution case, no closed-form expression for the rate achievable with finite-cardinality constellations is available)}.
The approximation relies on the following assumption: the real part $\hat{x}_k^R = \Re\{\hat{x}_k\}$ and the imaginary part $\hat{x}_k^I = \Im\{\hat{x}_k\}$ of the soft estimate $\hat{x}_k$ of the transmitted symbol $x_k$ are conditionally jointly Gaussian given $x_k$ and $\hat{\matH}$, with conditional mean $\vecmu(x_k, \hat{\matH})$ and conditional covariance $\matSigma(x_k, \hat{\matH})$.
We use this assumption to approximate~\eqref{eq:achRate_MassiveMIMO} as follows (see Appendix~\ref{app:derivation}):
\begin{IEEEeqnarray}{rCl}
R^{(k)}(\snr)
&\approx& {\frac{T-P}{T}} \bigg( h \lefto(\hat{x}_k^R, \, \hat{x}_k^I \given \hat{\matH}\right) \nonumber\\ 
&& - \frac{1}{2} \Ex{x_k, \hat{\matH}}{\log_2\left( \left( 2 \pi e\right)^2 \det{\matSigma(x_k, \hat{\matH})}\right)} \! \bigg). \IEEEeqnarraynumspace
\label{eq:highSNRapprox}
\end{IEEEeqnarray}
Here, {$h(\cdot\given \cdot)$ denotes the conditional differential entropy}~\cite{cover06a}.
Note that under the above Gaussian approximation, the conditional probability of $[\hat{x}_k^R, \, \hat{x}_k^I]^T$ given $\hat{\matH}$ is a Gaussian mixture, for which the differential entropy is not known in closed form but can be computed efficiently.
{The accuracy of the approximation in~\eqref{eq:highSNRapprox} depends crucially on the choice of $\vecmu(x_k, \hat{\matH})$ and $\matSigma(x_k, \hat{\matH})$.}
{In \fref{app:derivation}, we provide suitable choices for $\vecmu(x_k, \hat{\matH})$ and   $\matSigma(x_k, \hat{\matH})$ for the MRC case (see \eqref{eq:vecmuMRC}--\eqref{eq:Sigma21mrc}). 
For the ZF case, $\vecmu(x_k, \hat{\matH})$ is provided in \eqref{eq:vecmuMRC} whereas, to improve the accuracy of the approximation, we resort to  the numerical method described in~\fref{app:derivation} to compute $\matSigma(x_k, \hat{\matH})$.
As we shall illustrate in \fref{sec:one_bit_num}, the resulting approximation~\eqref{eq:highSNRapprox} turns out to be accurate for all system parameters considered in this paper.}

\subsection{Sum-Rate Approximation for Gaussian Inputs}
\label{sec:gaussapprox}
{Next, we present an approximation on the achievable rate~\eqref{eq:achRate_MassiveMIMO} assuming Gaussian inputs. 
In contrast to \cite{li17b, mollen16c}, where a similar approximation is derived for the 1-bit case, we shall consider the case of multi-bit ADCs. 

The approximation relies on Bussgang's decomposition and on the assumption that the quantizer input $\vecy$ can be modeled as a  Gaussian random vector\footnote{To keep notation compact, we drop in this section the time index $t$ since it is superfluous.}  and that its covariance matrix satisfies  
\begin{equation}\label{eq:cov_matrix_rate}
  \matC_{\vecy} = (K\snr + 1)\matI_N.
\end{equation}
Both the Gaussian assumption and~\eqref{eq:cov_matrix_rate} are accurate at low SNR or when the number of UEs is large.
Under these assumptions, we can use Bussgang's theorem to decompose the received signal as%
\begin{IEEEeqnarray}{rCl}
\vecr &=&Q_b(\vecy)= G_b\vecy + \vecd \label{eq:lololo}
\end{IEEEeqnarray}%
where $\vecd$ is the quantization distortion. 
Here, we have used that $\matG_b =G_b\matI_N$, which follows from~\eqref{eq:cov_matrix_rate}.
Furthermore, due to the power normalization~\eqref{eq:alpha} and due to~\eqref{eq:cov_matrix_rate}, the covariance matrix $\matC_\vecr$ of $\vecr$ satisfies $\matC_\vecr=(K\rho+1)\matI_N$. 
Hence, the covariance matrix $\matC_\vecd$ of the quantization distortion $\vecd$ must be equal to 
\begin{IEEEeqnarray}{rCl}
\matC_{\vecd} = \matC_{\vecr}- G_b^2\matC_{\vecy}= \lefto(1 - G_b^2\right)\lefto(K \snr + 1 \right)\matI_N.
\end{IEEEeqnarray}
Substituting~\eqref{eq:lololo} into~\eqref{eq:linreceived}, we obtain
\begin{IEEEeqnarray}{rCl}
\hat{x}_{k} &=&  \veca_k^H(G_b\vecy + \vecd) = G_b \veca_k^H\hat\matH\vecx + \veca_k^H\vecn \label{eq:mf_output_gaussian}
\end{IEEEeqnarray}
where $\hat\matH$ is the channel estimate~\eqref{eq:LS_MIMO}, and $\tilde\matH$ is the corresponding estimation error.
Here, we have defined $\vecn = G_b\tilde\matH\vecx + G_b\vecw + \vecd$.
Note that the noise $\vecn$  and the input vector $\vecx$ are uncorrelated provided that both~\eqref{eq:cov_matrix_rate} and~\eqref{eq:approx_off_diag} hold. 
Assuming that this is indeed the case, we can approximate the mutual information~\eqref{eq:achRate_MassiveMIMO} by using the auxiliary channel lower bound~\cite[p.~3503]{arnold06a} and treating the additive noise $\veca_k^H\vecn$ in~\eqref{eq:mf_output_gaussian} as a Gaussian random variable.
Specifically, let 
\begin{IEEEeqnarray}{rCl} \label{eq:snr_effective}
\bar\snr &=& \frac{G_b^2\hat\sigma^2\snr}{G_b^2K\tilde\sigma^2\snr + G_b^2 + (1 - G_b^2)(K\snr + 1)}
\end{IEEEeqnarray}
where $\hat\sigma^2$ and $\tilde\sigma^2$ are given by \eqref{eq:sigmahat} and \eqref{eq:sigmatilde}, respectively.
In~\eqref{eq:snr_effective}, the three terms in the denominator correspond to the estimation error, the additive noise, and the quantization distortion, respectively.
Since the channel input $\vecx$ is Gaussian, using~\cite[p.~3503]{arnold06a} we obtain the following approximation:
\begin{IEEEeqnarray}{rCl}
I(x_k;\hat{x}_k | \hat\matH) &\approx& \Ex{}{\log_2\!\Bigg( \! 1  + \! \frac{ \bar\snr|\veca_k^H\hat\vech_k|^2}{\bar\snr\!\sum\limits_{j \neq k} \! |\veca_k^H\hat\vech_j|^2 \! + \! \vecnorm{\veca_k}^2}\Bigg) \!}\!\! . \IEEEeqnarraynumspace \label{eq:Igauss}
\end{IEEEeqnarray}
Under the additional assumption that $\hat\matH$ is Gaussian, we can use~\cite[Eq.~(16) and~(20)]{ngo13a} to further lower-bound~\eqref{eq:Igauss} and obtain the following closed-form Gaussian approximations for the rates achievable with MRC and ZF, respectively:
\begin{IEEEeqnarray}{rCl} \label{eq:rate_approx_MRC}
R_\text{MRC}(\bar{\snr}) &\approx& \frac{T-P}{T}\log_2\lefto( 1 + \frac{(N-1)\bar\snr}{(K-1)\bar\snr + 1}\right)
\end{IEEEeqnarray}
and
\begin{IEEEeqnarray}{rCl} \label{eq:rate_approx_ZF}
R_\text{ZF}(\bar{\snr}) &\approx& \frac{T-P}{T}\log_2\lefto( 1 + (N-K)\bar\snr\right).
\end{IEEEeqnarray}
Here, we have multiplied the log terms by~$(T-P)/T$ to take into account the pilot overhead. 
Note that for the infinite-resolution case~($G_\infty = 1$), we recover from~\eqref{eq:rate_approx_MRC} and~\eqref{eq:rate_approx_ZF} the achievable rate with imperfect CSI reported in \cite[Eq.~(39)]{ngo13a} and \cite[Eq.~(42)]{ngo13a} for the MRC and ZF receiver, respectively. For the case of 1-bit ADCs ($G_1 = \sqrt{2/\pi}$), we recover from~\eqref{eq:rate_approx_ZF} the achievable rate approximation with ZF recently reported in \cite{li17b}. 

As we shall demonstrate in \fref{sec:one_bit_num}, despite the several assumptions invoked to obtain \eqref{eq:rate_approx_MRC} and \eqref{eq:rate_approx_ZF}, these approximations turn out to be accurate in the low-SNR regime.
}

\section{Numerical Results}
\label{sec:one_bit_num}

We now assess the rates achievable with the above detailed channel estimation and data-detection schemes detailed in the previous section on a massive MU-MIMO uplink system where the receiver is equipped with low-resolution ADCs.
 We assume that the users are able to coordinate the transmission of their pilots: when one of the UEs transmits pilots, the other UEs remain idle. In other words, pilots are transmitted in a round robin fashion.\footnote{This pilot-transmission method is chosen for convenience; it may be suboptimal.}
The use of time-interleaved pilots ensures that~$\sum_{t=1}^P \vecx_t\vecx_t^H = P\snr\, \matI_K$.
Also, because of the idle time, each user can transmit its pilots at a power level that is $K$ times higher than the power level for the data symbols, while still satisfying the average-power constraint~\eqref{eq:avgpow}.

\subsection{Channel Estimation}

\begin{figure}[!t]
	\captionsetup[subfloat]{farskip=0pt}
	\subfloat[$P = K$ pilots.]{
		\includegraphics[width=.5\figwidth]{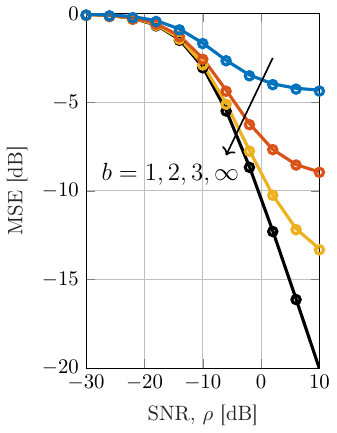}
		\label{fig:mse_1pilot}
	}
	\subfloat[$P = 3K$ pilots.]{
		\includegraphics[width=.5\figwidth]{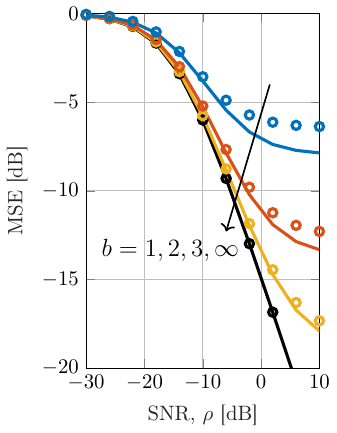}
		\label{fig:mse_3pilot}
	}
	\caption{MSE of the channel estimator~\eqref{eq:LS_MIMO} as a function of the SNR $\snr$; QPSK pilots, $N = 200$, $K = 10$. The solid lines correspond to the MSE approximation \eqref{eq:sigmatilde} and the marks correspond to the exact MSE, which was computed numerically.}
	\label{fig:mse}
\end{figure}

We start by validating the accuracy of the approximation for the MSE of the simplified channel estimator \eqref{eq:LS_MIMO} given in \eqref{eq:sigmatilde}.
Specifically, we compare in \fref{fig:mse} the exact MSE of the estimator \eqref{eq:LS_MIMO}, which is evaluated numerically, with the approximation \eqref{eq:sigmatilde}, for different values of SNR $\snr$, number of pilots $P$, and ADC resolution~$b$.

We note that if $P=K$ (\fref{fig:mse_1pilot}), then the MSE approximation \eqref{eq:sigmatilde} is indeed exact (as we claimed in \fref{sec:channel_est}). For the case of $P = 3K$ (\fref{fig:mse_3pilot}), the approximation \eqref{eq:sigmatilde} turns out to be accurate at low SNR. 
Furthermore, the accuracy of~\eqref{eq:sigmatilde} increases with the resolution of the ADCs. Indeed,~\eqref{eq:sigmatilde} relies on the assumption that the off-diagonal elements of the covariance matrix in~\eqref{eq:approx_off_diag} are zero and these entries vanish as the ADC resolution increases~(see, e.g., \cite[p.~541]{widrow08a} for more details).

\subsection{Achievable Rate}

\subsubsection{Single-User Case, 1-bit ADCs}

\begin{figure}[!t]
	\centering
	\includegraphics[width=\figwidth]{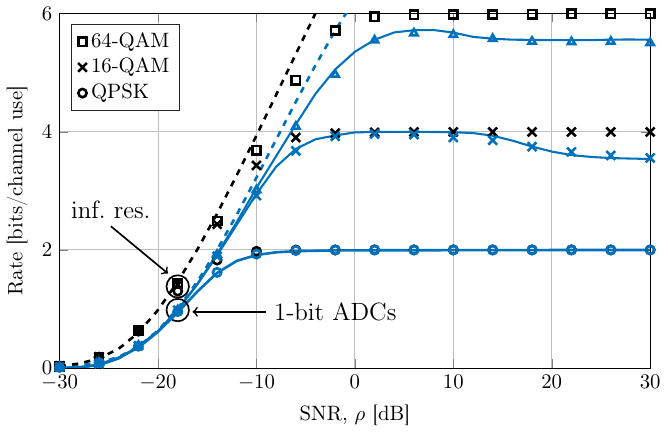}
	\caption{Single-user achievable rate with MRC as a function of the SNR $\snr$; $N = 200$, $K=1$, $T=1142$; the number of pilots~$P$ is optimized for each value of $\snr$. The solid lines correspond to the finite-cardinality approximation~\eqref{eq:highSNRapprox}, the dashed lines corresponds to the Gaussian approximation \eqref{eq:rate_approx_MRC}, and the marks correspond to the rates computed via \eqref{eq:achRate_MassiveMIMO} and \eqref{eq:mutinfo_MassiveMIMO}.}
	\label{fig:single_user_rate_vs_snr_1bit}
\end{figure} 

In \fref{fig:single_user_rate_vs_snr_1bit} we compare for the single-user 1-bit ADC case, the rates achievable with QPSK, 16-QAM, and 64-QAM as a function of $\snr$ for the MRC receiver.\footnote{To evaluate the mutual information~\eqref{eq:mutinfo_MassiveMIMO}, we have simulated $300$ random fading channel realizations. For each channel realization we have considered $3000$ random noise realizations for each symbol in the constellation.}
We depict both the rates achievable with 1-bit ADCs and the ones for the infinite-resolution case. The rates with 1-bit ADCs, which are computed using \eqref{eq:achRate_MassiveMIMO} and \eqref{eq:mutinfo_MassiveMIMO}, are compared with the approximation for finite-cardinality constellations provided in \eqref{eq:highSNRapprox} {and the Gaussian approximation~\eqref{eq:rate_approx_MRC}} to verify their accuracy.
The infinite-resolution rates are computed using \eqref{eq:achRate_MassiveMIMO} and \eqref{eq:mutinfo_MassiveMIMO} {and are also compared with the Gaussian approximation~\eqref{eq:rate_approx_MRC}}
The number of receive antennas is $N = 200$ and the coherence interval is $T=1142$.\footnote{For an LTE-like system operating at $2$ GHz, with symbol time equal to $66.7$\,$\upmu$s, and with UEs moving at a speed of $3$\,km/h, the duration of the coherence interval according to Jake's model is approximately $T=1142$~symbols.}
The number of transmitted pilots $P$ is numerically optimized for every value of $\snr$.
We see that, despite using 1-bit ADCs, higher-order modulations outperform QPSK already at SNR values as low~as~$\snr = -15$\,dB.

Note that the achievable rate does not increase monotonically with $\snr$ in the 16-QAM and 64-QAM case. Indeed, as $\snr$ gets large the constellation gets projected onto the unit circle and the number of distinguishable constellation points becomes smaller (see \fref{fig:constc}).
Note also that the approximation \eqref{eq:highSNRapprox} for finite-cardinality constellations closely tracks the simulation results {for all SNR values}.
This approximation enables us to accurately predict the SNR value beyond which the rates achievable with a given constellation saturates. This, in turn, allows us to identify the most appropriate constellation for a given SNR value.

The Gaussian approximation~\eqref{eq:rate_approx_MRC} tracks the rates achievable with finite-cardinality constellations accurately in the low-SNR~regime.

We note that, when QPSK is used, the difference in the achievable rates between the 1-bit quantized case and the infinite-resolution case is marginal---an observation that was already reported in~\cite{risi14a}. 
In contrast, the rate loss is more pronounced for higher-order constellations.

\subsubsection{Multi-User Case, 1-bit ADCs}

\begin{figure}[!t]
	\centering
	\captionsetup[subfloat]{farskip=0pt}
	\subfloat[MRC receiver.]{
		\includegraphics[width=.5\figwidth]{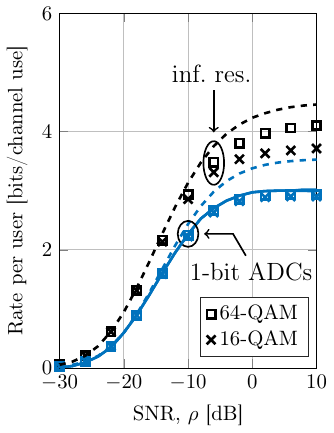}
	} 
	\subfloat[ZF receiver.]{
		\includegraphics[width=.5\figwidth]{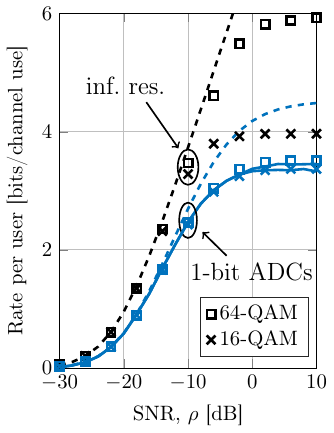}
		\label{fig:multi_user_rate_vs_snr_1bit_zf}
	}
	\caption{Per-user achievable rate as a function of the SNR $\snr$; $N = 200$, $K=10$, $T=1142$; the number of pilots $P$ is optimized for each value of $\snr$. The solid lines correspond to the finite-cardinality approximation \eqref{eq:highSNRapprox}, the dashed lines corresponds to the Gaussian approximations \eqref{eq:rate_approx_MRC}, \eqref{eq:rate_approx_ZF}, and the marks correspond to the rates computed via \eqref{eq:achRate_MassiveMIMO} and \eqref{eq:mutinfo_MassiveMIMO}.}
	\label{fig:multi_user_rate_vs_snr_1bit}
\end{figure}

In \fref{fig:multi_user_rate_vs_snr_1bit}, we plot the rates achievable with MRC and ZF for both the 1-bit-ADC and the infinite-resolution case when $K=10$ users are active. 
Motivated by the results in \fref{fig:single_user_rate_vs_snr_1bit}, we only compare the rates achievable with 16-QAM and 64-QAM.
Note again that the approximation \eqref{eq:highSNRapprox} turns out to be accurate for a {all SNR values, whereas the Gaussian approximation is accurate only at low SNR. Note also that rates with 16-QAM and 64-QAM saturate at the same level at high SNR for both MRC and ZF. This implies that the system is effectively distortion and interference limited, and that the Gaussian approximations~\eqref{eq:rate_approx_MRC}, \eqref{eq:rate_approx_ZF} overestimate the rate  for high SNR values.}

\begin{figure}[!t]
	\centering
	\includegraphics[width=\figwidth]{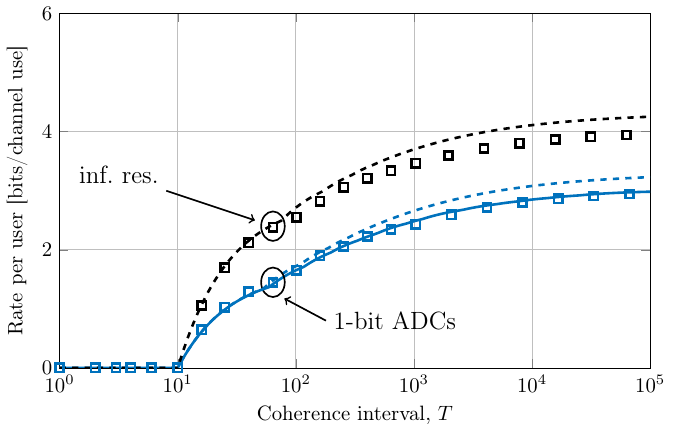}
	\caption{Per-user achievable rate with 64-QAM and ZF as a function of $T$; $\snr = -10$\,dB, $N = 200$, $K=10$; the number of pilots $P$ is optimized for each value of $T$. The solid lines correspond to the finite-cardinality approximation \eqref{eq:highSNRapprox}, the dashed lines corresponds to the Gaussian approximation~\eqref{eq:rate_approx_ZF}, and the marks correspond to the rates computed via \eqref{eq:achRate_MassiveMIMO} and \eqref{eq:mutinfo_MassiveMIMO}.}
	\label{fig:multi_user_rate_vs_T_1bit}
\end{figure}

\subsubsection{Dependence on the Coherence Interval}

In \fref{fig:multi_user_rate_vs_T_1bit}, we plot the per-user achievable rates with ZF, as a function of the coherence interval~$T$ for $\snr = -10$\,dB, $N = 200$, $K = 10$, and 64-QAM constellation. 
{We observe that the reduction in the achievable rate when $T$ is made smaller is similar for both the 1-bit and infinite-resolution case. Hence, operating in a high-mobility scenario leads to similar performance losses in both cases.}
Note also that the achievable rate is zero when $T \le 10$. 
In fact, when orthogonal pilot sequences are transmitted, at least $10$ pilot symbols are required when~$K=10$.

\subsubsection{Dependence on ADC Resolution}

\begin{figure}[!t]
	\centering
	\includegraphics[width=\figwidth]{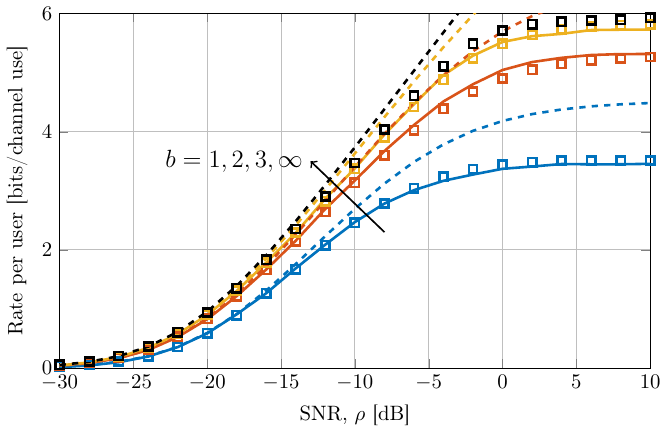}
	\caption{Per-user achievable rate with 64-QAM and ZF as a function of the SNR $\snr$; $N = 200$, $K=10$, $T=1142$; the number of pilots $P$ is optimized for each value of $\snr$. The solid lines correspond to the finite-cardinality approximation \eqref{eq:highSNRapprox}, the dashed lines corresponds to the Gaussian approximation~\eqref{eq:rate_approx_ZF}, and the marks correspond to the rates computed via~\eqref{eq:achRate_MassiveMIMO}~and~\eqref{eq:mutinfo_MassiveMIMO}.}
	\label{fig:multi_user_rate_vs_snr_both}
\end{figure}

Focusing on 64-QAM and ZF, we compare in \fref{fig:multi_user_rate_vs_snr_both} the achievable rate as a function of the ADC resolution and the SNR.
We observe that with 2-bit ADCs, the achievable rate increases significantly compared to the 1-bit-ADC case. 
For example, at $\snr = -10$\,dB, we achieve 90\% of the infinite-resolution rate, compared to 71\% with 1-bit ADCs.  
Increasing the ADC resolution beyond $3$~bits seems unnecessary for the system parameters considered in \fref{fig:multi_user_rate_vs_snr_both}. 
{This conclusion is supported by both the approximation for finite-cardinality constellations and the one for Gaussian inputs.
}{We note that the Gaussian approximation~\eqref{eq:rate_approx_ZF} is again accurate at low SNR. Furthermore, as expected its accuracy increases with the ADC resolution.}

\subsection{Impact of Large-Scale Fading and Imperfect Power Control}
\label{sec:large_scale}

So far, we have considered only the case when all users operate at the same average SNR. This corresponds to the scenario where perfect power control can be performed in the uplink, which is clearly favorable for low-resolution ADC architectures. 
If, however, the received signal powers are vastly different, low-power signals may not be distinguishable from high-power interferers for cases in which the ADCs resolution is too low.

In practical systems, large spreads in the received power is typically avoided through power control. 
However, perfect power control may be impossible to achieve in practice due to limitations on the UE transmit power, for example. 
We next investigate how relaxing the accuracy of the UE transmit power control will impact the system performance.
We consider a single-cell scenario and adapt the urban-macro path loss model in \cite{3gpp14a}.
The simulation parameters for this study are summarized in Table~\ref{tab:simulation_parameters}. The transmit power for all UEs is set to 8.5\,dBm, which for the first user that is located $d_1 = 185$ meters from the BS, results in a SNR of approximately $\snr_1 = -10$\,dB.
The remaining $K-1$ users in the cell are randomly dropped according to a uniform distribution on the circular ring of inner radius $d_1 - \Delta d$ meters and outer radius $d_1 + \Delta d$ meters, for a distance spread $0 < \Delta d < 150$~meters.
The case $\Delta d = 0$ corresponds to the scenario when power control is executed perfectly. The case $\Delta d = 150$ meters corresponds to the worst-case scenario of \emph{uncoordinated} uplink transmission, where no power control is performed by the UEs. In the latter case, the SNR for each interfering user lies in the range $[-19.0\,\text{dB},~15.3\,\text{dB}]$.

\begin{table}[!t]
	\makeatletter
	\renewcommand{\arraystretch}{.95}
	\makeatother
	\centering
	\caption{Summary of simulation parameters}
	\label{tab:simulation_parameters}
	\begin{tabular}{ll}
		\toprule
		\small Description & \small Assumption \\ 
		\midrule
		\normalsize
		\small Cell layout & \small Circular cell \\    
		\small Cell radius & \small 335\,meters \\    
		\small Minimum distance between UE and BS & \small 35\,meters \\
		\small Path loss & \small $35 + 35\log_{10}(d)$\,dB \\
		\small Number of BS antennas ($N$) & \small 200 antennas \\    
		\small Number of single-antenna users \small ($K$) & 10 users\\    
		\small Coherence interval ($T$) & \small 1142 channel uses \\    
		\small Number of pilots per user ($P/K$) & \small 10 pilots per user \\
		\small Carrier frequency & \small 2\,GHz \\
		\small System bandwidth & \small 20\,MHz \\
		\small UE transmit power & \small 8.5\,dBm \\
		\small Noise spectral efficiency &  \small $-$174.2\,dBm/Hz \\  
		\small Noise figure & \small 5\,dB \\ 
		\bottomrule
	\end{tabular}
\end{table}

\begin{figure}[t!]
	\centering
	\subfloat[MRC receiver.]{
	\includegraphics[width=.51\figwidth]{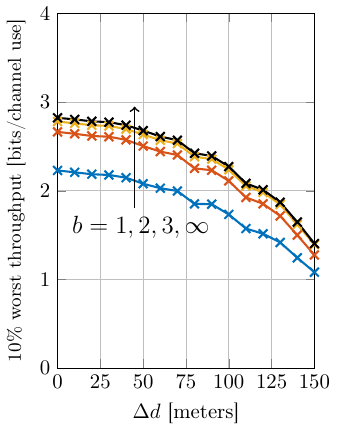}}
	\subfloat[ZF receiver.]{
	\includegraphics[width=.51\figwidth]{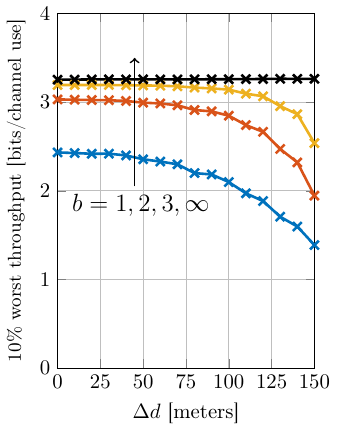}}
	\caption{The 10\% worst throughput  with 16-QAM for a user located $d_1 = 185$ meters away from the BS as a function of $\Delta d$ for the parameters specified in~Table~\ref{tab:simulation_parameters}.}
	\label{fig:multi_user_rate_vs_dist_spread}
\end{figure} 

In \fref{fig:multi_user_rate_vs_dist_spread}, we plot the 10\% worst throughput (i.e., the throughput corresponding to the 10\% point of the CDF of throughputs), for the intended user located $d_1 =185$ meters away from the BS, as a function of $\Delta d$. We focus on 16-QAM and assume that the received signal power level for each user is known to the BS.
To attain the curves, we have considered $10^3$ random interfering user drops for each $\Delta d$ value.
As expected, the gap to the infinite-resolution rate grows as $\Delta d$ increases.
In the uncoordinated case, with 1-bit ADCs and ZF, we attain~$57$\% of the rate achievable with perfect power control. The corresponding number for the $3$-bit-ADC case is~$79$\%.
This shows that high rates are achievable with low-resolution ADCs even in absence of power control.

\section{Conclusions}
\label{sec:conclusions}

We have analyzed the performance of a low-resolution quantized uplink massive MU-MIMO system operating over a frequency flat Rayleigh block-fading channel whose realizations are not known \emph{a priori} to transmitter and receiver.
In particular, we have shown that for the 1-bit massive MIMO case, high-order constellations, such as 16-QAM, can be used to convey information at higher rates than with QPSK; this holds in spite of the nonlinearity introduced by the 1-bit ADCs.
Furthermore, reliable communication can be achieved by using simple signal processing techniques at the receiver, i.e., {pilot-based channel estimation based on the Bussgang decomposition~\eqref{eq:LS_MIMO}} and MRC~detection.
By increasing the resolution of the ADCs by only a few bits, e.g., to 3 bits, we can achieve near infinite-resolution performance for a broad range of system parameters; furthermore, the system becomes robust against differences in the received signal power from the different users, due for example, to large-scale fading or imperfect power~control.

An extension of our analysis to a OFDM based setup for transmission over frequency-selective channels is currently under investigation. Such an extension could be used to benchmark the results recently reported in \cite{studer16a} in which the authors reported that, with a specific modulation and coding scheme taken from IEEE 802.11n, $4$ to $6$ bits are required to achieve a packet error rate below $10^{-2}$ at an SNR close to the one needed in the infinite-resolution case. 
We conclude that for a fair comparison between the performance attainable using low-resolution versus high-resolution ADCs, one should take into account the overall power consumption, including the power consumed by RF and baseband processing circuitry.

\appendices

\section{Proof of \fref{thm:general}}
\label{app:appA}
{It follows from Bussgang's theorem~\cite{bussgang52a} that
}%
{\begin{IEEEeqnarray}{rCl} \label{eq:mandag_hela_veckan}
  \Ex{}{\vecr\vecy^H} = \matG_b\Ex{}{\vecy\vecy^H}
\end{IEEEeqnarray}
}%
{where $\matG_b$ is a $N \times N$ diagonal matrix with 
}%
{\begin{IEEEeqnarray}{rCl} \label{eq:Gbb}
[\matG_b]_{n.n}  &=& \frac{1}{\sigma_n^2}\Ex{}{Q_b(y_n)y_n^*}.
\end{IEEEeqnarray}
}%
{Here, $y_n$ denotes the $n$th entry of the vector $\vecy$, $n=1,\dots,N$, and $\sigma_n^2 = \Ex{}{\abs{y_n}^2} = [\matK]_{n,n}$. 
It follows from \eqref{eq:mandag_hela_veckan} that we can write the quantized signal as $\vecr = \matG_b\vecy + \vecd$, where $\vecd$ and $\vecy$ are uncorrelated.
}%
{Note now that the quantizer output $Q_b(y_n)$  is equal to $\ell_i + j\ell_i$ if and only if $\Re\{y_n\} \in [\tau_{i}, \tau_{i+1})$ and $\Im\{y_n\} \in [\tau_{j}, \tau_{j+1})$. Thus, 
}%
{\begin{IEEEeqnarray}{rCl} 
\Ex{y_n}{Q_b(y_n)y_n^*} 
&=& \!\!\int_{-\infty}^\infty  \frac{2 Q_b(z) z }{\sigma_n\sqrt{\pi}} \exp\lefto( \!- \frac{z^2}{\sigma_n^2}\right)\! \mathrm{d}z \\
&=& \sum_{i = 0}^{L-1}  \frac{\ell_i\sigma_n}{\sqrt{\pi}}  \lefto( e^{-\frac{\tau_i^2}{\sigma_n^2}} - e^{-\frac{\tau_{i+1}^2}{\sigma_n^2}} \right) \!. \IEEEeqnarraynumspace \label{eq:expandexp}
\end{IEEEeqnarray}
}%
{We obtain~\eqref{eq:gainmatrix_general} by substituting~\eqref{eq:expandexp} in~\eqref{eq:Gbb} and by using that~$\sigma_n^2 = [\matK]_{n,n}$.
}

\section{Derivation of \eqref{eq:highSNRapprox}}
\label{app:derivation}

To keep the notation compact, we set $\hat{x}_k^R = \Re\{\hat{x}_k\}$ and $\hat{x}_k^I = \Im\{\hat{x}_k\}$. By letting $a_{n,k} = a_{n,k}^R + ja_{n,k}^I$, $h_{n,k} = h_{n,k}^R + jh_{n,k}^I$, and $r_{n} = r_{n}^R + jr_{n}^I$, where $a_{n,k}$ denotes the $n$th entry of the receive filter $\veca_k$ and $r_n$ the $n$th entry of the received vector $\vecr$, we can express the real components of the received signal as
\begin{IEEEeqnarray}{rCl}
\hat{x}_k^R
&=& \sum_{n=1}^N \Re\{ a_{n,k}^* r_n \} =  \sum_{n=1}^N \lefto( a_{n,k}^R r_n^R +  a_{n,k}^I\, r_n^I \right).
\label{eq:real_received}
\end{IEEEeqnarray}
Similarly, for the imaginary part we can write
\begin{IEEEeqnarray}{rCl}
\hat{x}_k^I
&=& \sum_{n=1}^N \Im\{ a_{n,k}^* r_n \} =  \sum_{n=1}^N \lefto( a_{n,k}^R r_n^I -  a_{n,k}^I\, r_n^R \right).
\label{eq:imag_received}
\end{IEEEeqnarray}
Now, we collect the real and imaginary components in a vector $[\hat{x}_k^R, \, \hat{x}_k^I]^T$ and approximate their conditional distribution given the channel input and the channel estimate as a bivariate Gaussian random vector with mean $\vecmu(x_k, \hat{\matH})$ and $2 \times 2$ covariance matrix $\matSigma(x_k, \hat{\matH})$. Under this assumption, we have~that
\begin{IEEEeqnarray}{rCl}
I(x_k; \hat{x}_k \given \hat{\matH}) 
&=& h\lefto(\hat{x}_k^R, \, \hat{x}_k^I \given \hat{\matH}\right) \nonumber\\
&& - \frac{1}{2} \Ex{x_k, \hat{\matH}}{\log_2\lefto( \! \left( 2 \pi e\right)^2 \! \det{\matSigma(x_k, \hat{\matH})} \!\right)}\!. \IEEEeqnarraynumspace
\label{eq:mutinfoexpansion}
\end{IEEEeqnarray}
{It is worth emphasizing  that, differently from~\eqref{eq:mutinfo_MassiveMIMO}, the differential entropy $h(\hat{x}_k^R, \, \hat{x}_k^I \given \hat{\matH})$ in \eqref{eq:mutinfoexpansion} is evaluated under the assumption that $[\hat{x}_k^R, \, \hat{x}_k^I]^T$ is conditionally Gaussian given~$x_k$ and~$\hat{\matH}$.} The conditional probability of $[\hat{x}_k^R, \, \hat{x}_k^I]^T$ given~$\hat{\matH}$ is a Gaussian mixture.
The achievable rate in~\eqref{eq:highSNRapprox} follows from~\eqref{eq:mutinfoexpansion} by taking into account the rate loss due to the transmission of $P$ pilot symbols to estimate the channel. 
We shall next discuss how to choose $\vecmu(x_k, \hat{\matH})$ and $\matSigma(x_k, \hat{\matH})$.

{We start by finding a suitable approximation for the probability mass functions of the random variables $r_n^R$ and $r_n^I$. For $r_n^R$, it holds that 
\begin{IEEEeqnarray}{rCl}
p^R_{n,i} &=& \Pr\lefto\{r_n^R= \ell_i \right\} \\
&=& \Pr\left\{ r_n^R < \tau_{i+1}\right\} - \Pr\left\{ r_n^R \le \tau_i \right\} \\
&\approx& \Phi(\zeta_{i+1}^R) - \Phi(\zeta_{i}^R) \IEEEeqnarraynumspace \label{eq:someprob}
\end{IEEEeqnarray}
where in the last step we have approximated the interference term $\sum_{j \neq k} \Re\{ h_{n,j}x_j \}$ by a zero-mean Gaussian random variable with variance $\snr \sum_{j \neq k} \abs{h_{n,j}}^2$ and defined
\begin{IEEEeqnarray}{rCl}
\zeta_i^R &=& \sqrt{ \frac{2\lefto(\tau_i - h_{n,k}^R x_{k}^R + h_{n,k}^I x_{k}^I\right)}{1 + \snr \sum_{j \neq k} \abs{h_{n,j}}^2}} .
\end{IEEEeqnarray}
For the single-user case, the approximation \eqref{eq:someprob} is exact since there is no interference. 
Proceeding in an analogous way, we can show that 
\begin{IEEEeqnarray}{rCl}
p^I_{n,i} &\approx& \Phi(\zeta_{i+1}^I) - \Phi(\zeta_{i}^I)
\end{IEEEeqnarray}
where
\begin{IEEEeqnarray}{rCl}
\zeta_i^I = \sqrt{ \frac{2\lefto(\tau_i - h_{n,k}^R x_{k}^I - h_{n,k}^I x_{k}^R\right)}{1 + \snr \sum_{j \neq k} \abs{h_{n,j}}^2}} .
\end{IEEEeqnarray}
Next, we use these approximations to derive $\vecmu(x_k, \hat{\matH})$ and $\matSigma(x_k, \hat{\matH})$.
The conditional mean of $x_k^R$ given both $x_k$ and $\hat\matH$ can be written as
\begin{IEEEeqnarray}{rCl}
\Ex{}{\hat{x}_k^R \given x_k, \hat{\matH}}
&=& \sum_{n=1}^N \Ex{}{a_{n,k}^R r_n^R +  a_{n,k}^I\, r_n^I } =  \sum_{n=1}^N  m_{n,k}^R \IEEEeqnarraynumspace
\end{IEEEeqnarray}
where $m_{n,k}^R = \sum_{i=0}^{2^b-1} \ell_i \lefto( a_{n,k}^R p_{n,i}^R +  a_{n,k}^I p_{n,i}^I \right)$. 
Similarly, for the imaginary part it holds that
\begin{IEEEeqnarray}{rCl}
\Ex{}{\hat{x}_k^I \given x_k, \hat{\matH}}
&=&  \sum_{n=1}^N m_{n,k}^I
\end{IEEEeqnarray}
where $m_{n,k}^I = \sum_{i=0}^{2^b-1}\ell_i \lefto( a_{n,k}^R p_{n,i}^I -  a_{n,k}^I p_{n,i}^R \right)$. The sought-after mean vector can thus be written as
\begin{IEEEeqnarray}{rCl}
\vecmu(x_k, \hat\matH) 
&=& \sum_{n=1}^N \sum_{i=0}^{2^b-1}\ell_i
\begin{bmatrix}
a_{n,k}^Rp_{n,i}^R +  a_{n,k}^Ip_{n,i}^I   \\
a_{n,k}^Rp_{n,i}^I -  a_{n,k}^Ip_{n,i}^R
\end{bmatrix}.
\label{eq:vecmuMRC}
\end{IEEEeqnarray}

We next move to $\matSigma(x_k, \matH)$. Assuming that the received signal is conditionally uncorrelated over the antenna array, we obtain that
\begin{IEEEeqnarray}{rCl}
\left[\matSigma(x_k, \hat\matH)\right]_{1,1} 
&=&  \Ex{}{\lefto(\hat{x}_k^R - \mu_k^R \right)^2 \given x_k, \hat{\matH}} \\
&=& \sum_{n=1}^N  \sum_{i=0}^{2^b-1} \sum_{j=0}^{2^b-1} p_{n,i}^R p_{n,j}^I \nonumber \\ 
&& \lefto( a_{n,k}^R \ell_i + a_{n,k}^I \ell_j - m_{n,k}^R \right)^2. \IEEEeqnarraynumspace
\label{eq:Sigma11mrc}
\end{IEEEeqnarray}
Analogously, it holds that
\begin{IEEEeqnarray}{rCl}
\left[\matSigma(x_k, \matH)\right]_{2,2} 
&=&  \Ex{}{\lefto(\hat{x}_k^I - \mu_k^I \right)^2 \given x_k, \hat{\matH}} \\
&=& \sum_{n=1}^N  \sum_{i=0}^{2^b-1} \sum_{j=0}^{2^b-1} p_{n,i}^R p_{n,j}^I \nonumber\\
&&\lefto( a_{n,k}^R \ell_j - a_{n,k}^I \ell_i - m_{n,k}^I \right)^2. \IEEEeqnarraynumspace
\label{eq:Sigma22mrc}
\end{IEEEeqnarray}
Furthermore, 
\begin{IEEEeqnarray}{rCl}
\left[\matSigma(x_k, \matH)\right]_{1,2} 
&=& \Ex{}{\lefto(\hat{x}_k^R - \mu_k^R \right)\lefto(\hat{x}_k^I - \mu_k^I \right) \given x_k, \hat{\matH}} \nonumber\\ \IEEEeqnarraynumspace
&=& \sum_{n=1}^N  \sum_{i=0}^{2^b-1} \sum_{j=0}^{2^b-1} p_{n,i}^R p_{n,j}^I \nonumber\\
&&\lefto( a_{n,k}^R \ell_i + a_{n,k}^I \ell_j - m_{n,k}^R \right) \nonumber\\[5pt] 
&&\lefto( a_{n,k}^R \ell_j - a_{n,k}^I \ell_i - m_{n,k}^I \right). 
\label{eq:Sigma12mrc}
\end{IEEEeqnarray}
Finally, because of symmetry,
\begin{IEEEeqnarray}{rCl}
\left[\matSigma(x_k, \matH)\right]_{2,1} = \left[\matSigma(x_k, \matH)\right]_{1,2}.
\label{eq:Sigma21mrc}
\end{IEEEeqnarray}

For the ZF receiver, computing the covariance using~\eqref{eq:Sigma11mrc}--\eqref{eq:Sigma21mrc} does not yield a satisfactory approximation. Therefore, we resort to Monte-Carlo simulations to obtain the covariance.
Specifically, we find $\matSigma(x_k, \matH)$ by simulating several random noise and interference realizations for each point in the symbol constellation. 
Obtaining a sufficiently accurate estimate of $\matSigma(x_k, \matH)$ requires orders of magnitude fewer noise and interference realizations compared to estimating the probability mass functions in~\eqref{eq:mutinfo_MassiveMIMO}.
}

\section*{Acknowledgements}
The authors would like to thank Dr.~Fredrik Athley at Ericsson Research for fruitful discussions.


\bibliographystyle{IEEEtran}
\begin{spacing}{1}
\bibliography{IEEEabrv,confs-jrnls,publishers,svenbib}
\end{spacing}

\balance

\end{document}